\documentclass[twocolumn]{svjour3}          
\smartqed  
\usepackage{graphicx}
\usepackage[dvipsnames]{xcolor}

\usepackage{booktabs} 
\usepackage{amsmath}
\usepackage{amssymb}
\usepackage{amsfonts}
\usepackage{braket}
\usepackage{subcaption}
\usepackage{siunitx}
\usepackage{multirow}
\usepackage[ruled,linesnumbered]{algorithm2e}
\usepackage{algorithmicx}
\usepackage{soul}
\usepackage{varioref}
\usepackage{hyperref}

\SetCommentSty{mycommfont}

\begin{document}
\sloppy

\title{TIGER: Topology-aware Assignment using Ising machines}
\titlerunning{TIGER: Topology-aware Assignment using Ising machines}
\subtitle{Application to Classical Algorithm Tasks and Quantum Circuit Gates}

\author{Anastasiia Butko  \and 
Ilyas Turimbetov \and
George Michelogiannakis \and
David Donofrio \and
Didem Unat \and
John Shalf}

\authorrunning{A. Butko et al.}

\institute{A. Butko \and 
            G. Michelogiannakis \and 
            D. Donofrio \and 
            J. Shalf 
            \at Lawrence Berkeley National Laboratory \\ 
            Berkeley CA 94720, USA
            \email{\{abutko,mihelog,ddonofrio,jshalf\}@lbl.gov} \\
            \and
            I. Turimbetov \and D. Unat 
            \at Ko\c{c} University, Istanbul 34450, Turkey \\
            \email{\{iturimbetov18,dunat\}@ku.edu.tr}
}

\date{September 20, 2020}

\maketitle

\begin{abstract}
Optimally mapping a parallel application to compute and communication resources is increasingly important as both system size and heterogeneity increase. A similar mapping problem exists in gate-based quantum computing where the objective is to map tasks to gates in a topology-aware fashion. This is an NP-complete graph isomorphism problem, and existing task assignment approaches are either heuristic or based on physical optimization algorithms, providing different speed and solution quality trade-offs. Ising machines such as quantum and digital annealers have recently become available and offer an alternative hardware solution to solve this type of optimization problems. In this paper, we propose an algorithm that allows solving the topology-aware assignment problem using Ising machines. We demonstrate the algorithm on two use cases, i.e. classical task scheduling and quantum circuit gate scheduling. 
TIGER---topology-aware task/gate assignment mapper tool---implements our proposed algorithms and automatically integrates them into the quantum software environment. To address the limitations of physical solver, we propose and implement a domain-specific partition strategy that allows solving larger-scale problems and  a weight optimization algorithm that allows tuning Ising model parameters to achieve better restuls. We use D-Wave's quantum annealer to demonstrate our algorithm and evaluate the proposed tool flow in terms of performance, partition efficiency, and solution quality. Results show significant speed-up compared to classical solutions, better scalability, and higher solution quality when using TIGER together with the proposed partition method. It reduces the data movement cost by 68\% in average for quantum circuit assignment compared to the IBM QX optimizer~\cite{ibmq}.

\end{abstract}

\section{Introduction}

The task assignment problem aims to maximize application performance by balancing computational load among multiple and often heterogeneous processing units while reducing compute overhead. The task assignment problem has been shown to be equivalent to a graph isomorphism problem by Bokhari~\cite{Bokhari81}, which is known to be NP-complete~\cite{1676925,Hoefler:2011:GTM:1995896.1995909}. Therefore, many solvers for this problem are heuristic~\cite{843736} that inevitably tradeoff solution quality for computation speed, or physical optimization algorithms, such as simulated annealing~\cite{bollinger}, genetic techniques~\cite{951971}, and others. In addition, solvers can have different optimization metrics that are often contradictory, such as computational load, communication cost, or a weighted combination~\cite{6449795,7830488}. 

Scheduling quantum gates onto physical qubits is similarly a challenging problem, given the complexity and variety of quantum operations and physical restrictions of each quantum chip. To keep operations efficient, quantum gates should be scheduled on quantum hardware such as to minimize the number of operations and maximize quantum circuit fidelity (how much quantum information is preserved), while taking into account the connectivity between physical qubits~\cite{Guerreschi}. Consequently, many mapping algorithms scale poorly due to runtime, memory usage, and the quality of their generated solutions~\cite{2018arXiv180902573L}. In addition, the quality of their solutions compared to the theoretical optimal is unknown~\cite{2019arXiv190702026W}. These challenges indicate that gate assignment may hinder high-quality solutions on future quantum accelerators with more physical qubits and complex connectivity.

While genetic algorithms and simulated annealing are often considered best practices,
recent Ising machines offer an alternative hardware solution for a set of optimization problems, such as task scheduling. 
These Ising machines can be implemented using different technologies and exploit various physical effects. Such examples include coherent Ising machines \cite{optical}, Fujitsu's digital annealer \cite{fujitsu}, and quantum annealers designed by D-Wave Systems Inc. \cite{dwave}. Several studing on quantum annealers~\cite{2016PhRvA..94b2337M}~\cite{2017arXiv170104579K} explore its capabilities and limitations projecting the potential of these machines for future use.



Despite the potential benefits offered by quantum annealers combined with a growing interest in alternative solutions, practical applicability of annealing machines remains highly questionable. One of the reasons is physical limitations of current machines, namely the relatively small size of the chip and the poor connectivity between qubits \cite{2017arXiv170104579K}. Problem sizes demonstrated in comparison studies are usually not competitive with those handled by classical solvers. Therefore, effective problem partitioning and post-processing are required to continue exploiting quantum solver capabilities while the solution for physical limitations is sought \cite{Zick}. That makes most of the near-term quantum annealing-based approaches classical-quantum hybrids.

Another obstacle towards wide-spread quantum annealer adoption is programming complexity. Its programming model is based on the Quadratic Unconstrained Binary Optimization (QUBO)~\cite{Glover2018ATO} model that is different form the conventional programming and requires special approaches. 
The highest level that users are required to program D-Wave is ``virtual" QUBO, where ``virtual" means that the compiler takes care of mapping and routing the problem while taking into account device connectivity. Transforming a problem into QUBO format is not a trivial task. Higher-level tools as well as efficient algorithms are typically required \cite{10.1007/978-3-319-92040-5_6}.

{
In this work, we present the Topology-aware task assIGnment mappER (TIGER) to solve the assignment problem using Ising machines. Namely, our contributions are:
\begin{itemize}
    \item We develop an algorithm to assign Task-Communication Graph (TCG) to the architecture units minimizing the required data-movement and maximizing the performance. The assignment problem is expressed in the QUBO format to be used by an Ising machine.
    \item We develop an algorithm to assign Quantum Circuit Graph (QCG) to the qubits minimizing data movement (number of SWAP operations) and miximizing the fidelity. The assignment problem is expressed in the QUBO format to be used by an Ising machine.
    \item We develop a domain-specific QUBO partitioning algorithm (sub-QUBO) based on the graph dependency levels to overcome current physical limitations of existing quantum annealers and accelerate the solution search.
    \item We develop a weight optimization algorithm (WOA) to tune Ising equation parameters in order to prioritize target metrics and adjust them to obtain better solutions.
    \item We implement these algorithms as a TIGER tool. TIGER is written in Python and uses the NetworkX package~\cite{networkx} to create and manipulate TCG/QCG and ARC structures. 
    \item We integrate TIGER into the D-Wave tool-flow by supporting qbsolv qubo~\cite{qbsolv},  qmasm~\cite{7761637} formats and creating a feedback loop from D-Wave to TIGER in order to evaluate the solution for further optimizations.
    \item We evaluate the proposed algorithms and its implementation using D-Wave quantum annealer. We compare the D-Wave solver performance and quality of the task assignment (solution) to the classical TABU-search algorithm. We evaluate the quality of the quantum circuits assignment in terms of the circuit fidelity using real IBM systems~\cite{ibmq} and compare it against IBM QX gate optimizer. Our results show that TIGER with the D-Wave annealer provides up to 8\% of computation cost improvement and up to 25\% of communication cost improvement compared  to  the  classical  TABU-search solver when assigning a TCG. It reduces the data movement cost by 68\% in average for quantum circuit assignment compared to the IBM QX optimizer~\cite{ibmq}.
\end{itemize}
}

Given the relatively small size of the evaluated quantum annealer, we leave the discussion on general competitiveness of quantum annealers against classical computing out of the scope of this paper. Our results aim to provide useful insights on the entire tool-flow including classical decomposition, domain-specific partition and QUBO solvers. Last but not least, we would like to extend an invitation to the community to use TIGER and then contribute back to aid tool growth. Latest updates, documentation, and support can be found online \footnote{https://github.com/lbnlcomputerarch/tiger}.

The rest of the paper is organized as follows: Section \ref{sec:background} provides the background on the existing Ising machines. Section \ref{sec:map} and Section \ref{sec:qmap}  describe the proposed task assignment and quantum gate assignment mapping approaches, respectively. Section \ref{sec:tool} describes TIGER tool implementation as well as its integration into the complete tool-flow with the D-Wave programming environment. Section \ref{sec:res} shows performance, quality, sensitivity and scalability evaluation results. Section \ref{sec:con} concludes the work.
\section{Background}
\label{sec:background}



Ising machines are special-purpose processors that solve the Ising model, an intensely-studied NP-complete problem that is a system of interacting classical spins~\cite{Ising}.
An Ising model is mathematical model composed of a large lattice of sites, where each site can be in one of two states.
This model can be used to model the impact to the global state of the system caused by changes to parameters (such as connectivity and desired operations).
Ising models have been used to express and perform computation with different materials such as lasers and magnets, but are also the basis of several quantum accelerators because they are
a natural fit to express a graph of interconnected qubits.

\subsubsection{Quantum annealers}

Quantum annealing~\cite{Kadowaki98quantumannealing} is a metaheuristic technique for solving local search problems, such as finding the global minimum or maximum in a discrete search space. Quantum annealing offers potential benefits compared to popular heuristic algorithms through its quantum tunneling effect. This effect allows the system to penetrate energy barriers escaping from the local minima and therefore find better solutions to the original optimization problem. 

A quantum annealing machine or a quantum annealer is a hardware implementation of the adiabatic quantum computing algorithm. Quantum annealers operate on a set of qubits. A qubit is a two-state quantum-mechanical system that can carry states $\ket{0}$ and $\ket{1}$ or be in \textit{superposition} that expresses a linear superposition of the "basis states", i.e. $\ket{0}$ and $\ket{1}$. This feature forms the key power of quantum machines, which with $n$ qubits can be in an arbitrary superposition of up to $2^n$ different states simultaneously.
Another inherent quantum property of qubits is \textit{quantum entanglement} where a group of qubits is coupled to each other in such a way that the state of each qubit cannot be perceived separately, but as a whole system state instead~\cite{Nielsen:2011:QCQ:1972505}.

\begin{figure*}[t]
\includegraphics[width=\linewidth]{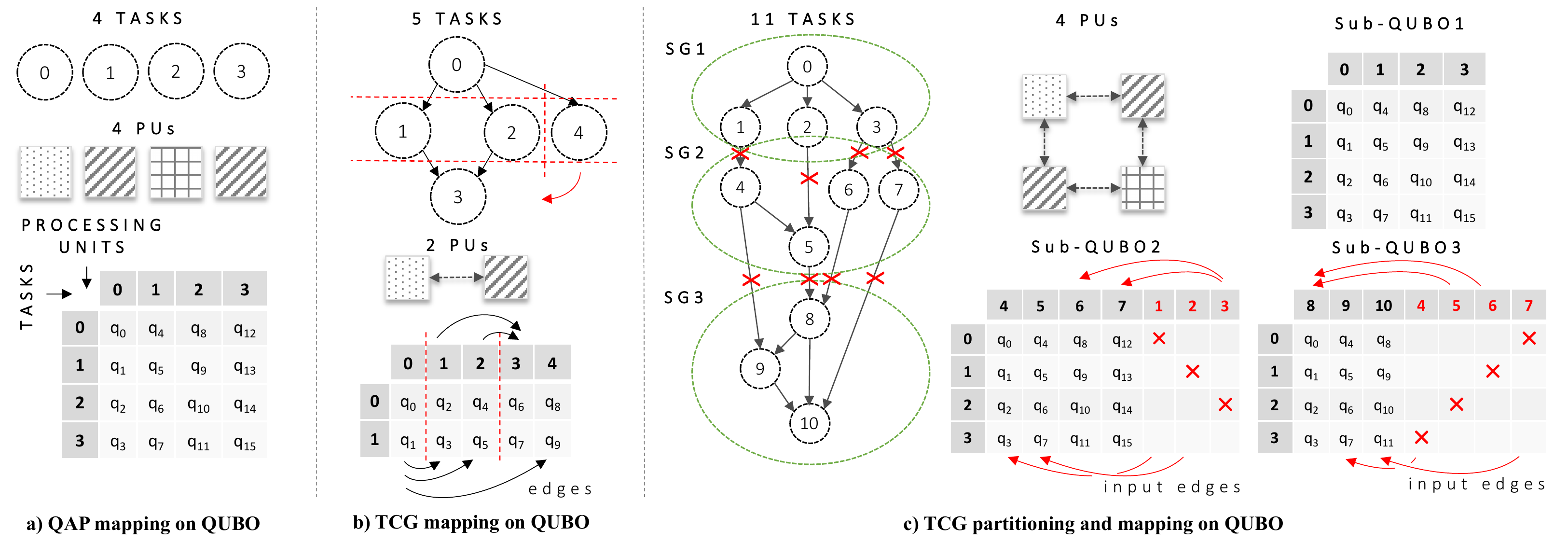}
\caption{Task Communication Graph (TCG) assignment on a heterogeneous multi-PU system: problem mapping on QUBO.}
\label{fig:map}
\end{figure*}

Quantum annealers provided by D-Wave Systems Inc. have been commercially available since 2011~\cite{dwave}. 
D-Wave quantum chips are implemented using superconducting technology and require an extreme isolated environment with a temperature close to absolute zero.
A closed cycle dilution refrigerator cools the processor down to 15 mK. Therefore, while the actual quantum chip is the size of a stamp, the physical volume of the whole D-Wave system reaches 20 $m^3$. However, D-Wave machines consume less than 25 kW of power, mostly for cooling and front-end servers~\cite{overview}.
In around 10 years, quantum annealing chips have reached $10^3$ number of qubits, promising significant performance improvement for certain computing problems in the near future. Physically, qubits are connected to each other using a so-called Chimera topology. The smallest Chimera unit contains a complete bipartite graph of eight vertices, each of which is connected to its four neighbours inside the unit and to its two neighbours outside the unit.

In~\cite{PhysRevX.6.031015}, authors compare the performance of physical quantum annealer (D-Wave 2X quantum annealer) to simulated annealing and quantum Monte Carlo methods executed on a classical processor.

Furthermore, authors in~\cite{2016PhRvA..94b2337M} extend Google Inc. studies by comparing quantum annealing to state-of-the art optimization methods, introducing more sophisticated assessment metrics. Their work considers four categories of optimization methods: sequential methods that include quantum annealing, simulated annealing and quantum Monte Carlo, tailored methods that solve simplified optimization problems, and non-tailored methods that are generic and thus represent the state of the art. Authors conclude that physical quantum annealing has better scaling compared to other sequential optimization methods, but it concedes to tailored as well as non-tailored state-of-the-art methods. Also, authors emphasize the importance of determining the application domain where quantum annealing maximizes its benefits, but this has yet to be defined.  

Finally, King et al. in~\cite{2017arXiv170104579K} introduce a problem class that can maximize usefulness of the quantum tunneling effect. Authors again compare quantum annealers to classical solvers and demonstrate three to four orders of magnitude performance speed-up in favor of quantum annealing.

Several studies demonstrate the use of quantum annealing for task scheduling. In \cite{schedRW}, authors introduce a hybrid quantum-classical approach to solving scheduling problems. Their framework integrates quantum annealing with classical computing into a guided tree search. Classical algorithms manage a global tree search and communicate the node search in QUBO format to the quantum annealer. Authors test the proposed framework on three scheduling problems, i.e. graph-coloring, mars lander task scheduling, and airport runway scheduling. Results show that the quantum annealer's output can effectively prune and guide the search process. Authors motivate their work by necessity to expand on the
capabilities of current quantum annealers and do not expect quantum annealers to be competitive in the near-term against classical computers.

In our work, we address a different scheduling problem, i.e. topology-aware assignment. The proposed TIGER framework extends existing software environments by automatically generating and dynamically adjusting QUBO files. We evaluate the tool flow in terms of quantum solver performance, the quality of task/gate assignment  and discuss the potential scalability of near-term machines.

\subsection{Problem formulation and programming}

Quantum annealers minimize the QUBO problem described by Equation~\ref{eq:1}. The equation describes the evolution of the time-dependent Hamiltonian \cite{Hwang:1987:HPL:2309685.2309803} that aims to find low-energy states in a system of $N$ interacting spins, i.e. qubits. In Equation~\ref{eq:1}, $q_i$ represents qubits that take value from the set $\{0, 1\}$, $h_i$ is a weight coefficient associated with each qubit, $J_{ij}$ denotes the strength of the couplings between two qubits $q_i$ and $q_j$ and $N$ is the number of qubits.

\begin{equation}
  E(q_1,...,q_N)=\sum_{i=1}^{N}h_{i}\cdot q_{i} + \sum_{i<j=1}^{N}J_{ij}\cdot q_i \cdot q_j
\label{eq:1}
\end{equation}

D-Wave annealer architectural designs impose a number of limitations on Equation~\ref{eq:1}. Notably, chips do not support all-to-all qubit connectivity. Thus, to couple two qubits located on different sides of the Chimera grid, excessive routing through other qubits is required. That dramatically cuts the number of available qubits to be purely used for problem solving. Another limitation concerns qubit weights and coupler strengths that lie in a specific range, i.e. [-2;2] and [-1;1] respectively, affecting the precision of the machine.


A low-level D-Wave program is expressed in the form of Equation~\ref{eq:1} as a list of $h_{i}$ and $J_{ij}$ with the associated qubit numbers. The provided solution is a list of $q_{i}$ values. This program is usually referred to as \textit{Quantum Machine Instruction} (QMI). At this level, all previously listed constrains, such as qubit connectivity, variable range as well as the number of physically available qubits, have to be taken into account. That makes D-Wave programming a challenging task. However, there are several tools to provide a certain level of abstraction by taking as input a so-called ``virtual'' QUBO that abstracts away the size or connectivity topology of the D-Wave system and maps the problem onto the physical hardware using different optimization techniques.

\begin{figure*}[t]
\includegraphics[width=\linewidth]{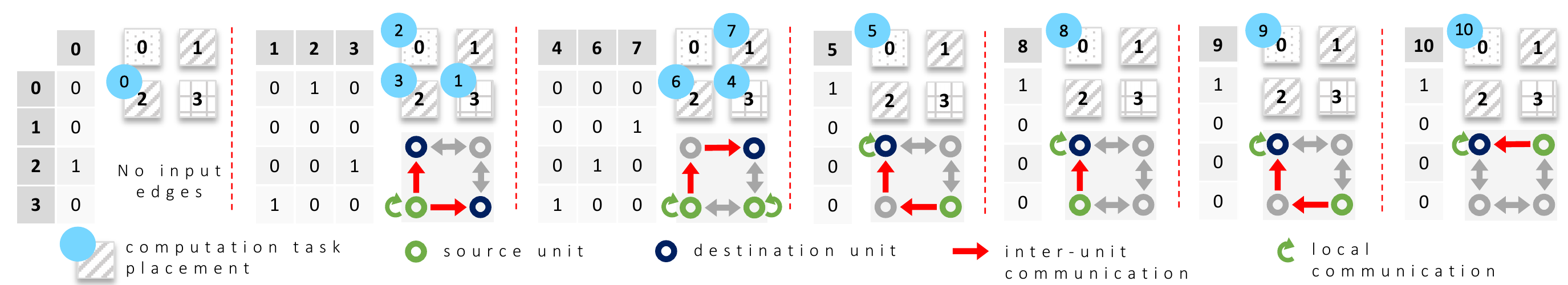}
\caption{Binary solution interpretation: computation task assignment and communication impact.}
\label{fig:map2}
\end{figure*}

\section{Task Assignment Mapping Algorithm}
\label{sec:map}

\subsection{Linear assignment problem}
In the task allocation context, the Linear Assignment Problem (LAP) consists of placing a set of independent tasks onto a set of Processing Units (PUs), with each assignment incurring a certain cost. The objective is to assign each task to a PU such that the total cost is minimized \cite{Burkard:2009:AP:1508120}. 

Figure~\ref{fig:map}(a) illustrates the transformation of the LAP to QUBO. The qubit matrix $Q$ represents the permutation matrix $X$, where each qubit defines the assignment of a task to a specific PU similar to $x_{ij}$ above. An $x_{ij}$ value of 1 represents that task $i$ was assigned to PU $j$.
A weight coefficient $h_i$ (not shown) represents the computational cost of the assignment. Since solvers in current machines find local minima, we transform positive computation costs into negative numbers to prevent the solver from giving all-zero answers.
To respect assignment constraints such as assigning one task to one qubit, we use qubit couplings and give them high penalty values such as $J_{ij} >> |h_i|$. For example, to prevent \textit{task 0} from being placed on multiple PUs, we couple qubits $(q_0 \cdot q_1)$, $(q_0 \cdot q_2)$, $(q_0 \cdot q_3)$, $(q_1 \cdot q_2)$, $(q_1 \cdot q_3)$ and $(q_2 \cdot q_3)$ for four qubits. Therefore, if two of these qubits have the same task assigned to them, the large penalty value will make the overall solution ineligible.

\subsection{Task-communication graph assignment}

Applications can be represented as a weighted directed acyclic graph, usually referred to as a Task Communication Graph (TCG). A TCG is defined as a tuple $G=(V, E)$, where $V=(v_i)$ is a set of weighted vertices with the weight representing task computational cost, and $E=(e_{i, j})$ is a set of weighted edges with the weight representing inter-task communication cost. 
An example of TCG is shown in the upper part of Figure~\ref{fig:map}(b).

Mapping of such as TCG into QUBO differs from previously shown LAP in three aspects. First, a TCG includes not only computation cost, but also inter-task communication cost expressed with graph edges. Second, not all tasks are assigned to PUs within the same time frame. A TCG is divided into multiple \textit{dependency levels} each of which represents a LAP. Dependency levels (groups) are shown with red dashed lines. Third, within each dependency level, the number of independent tasks can be different compared to the number of available PUs. The QUBO mapping transformation respects each of the above three constraints.

\textit{Communication edges.} Each communication edge is included into QUBO by qubit coupling. Communication cost is represented by coupling strength. Total end-to-end cost is calculated based on the weight of each edge in the communication path. If both source and destination tasks are assigned to the same PU, communication cost is equal to zero. This the most favourable case if the objective is to minimize data movement. For the example in Figure~\ref{fig:map}(b), to define the edge between \textit{task0} and \textit{task1} we couple qubits $(q_0 \cdot q_3)$ and $(q_1 \cdot q_2)$ with the associated topology-aware communication cost and qubits $(q_0 \cdot q_2)$ and $(q_1 \cdot q_3)$ with zero communication cost. 
Here, cost values are converted to negative numbers similar to computation cost values.
The relative priority of communication and computation costs can be formulated by adding a weight factor to bias the solver.

\textit{Dependency levels.}
Because of dependencies, only a certain number of tasks can be assigned to PUs in parallel. This relaxes the second assignment constraint that says that no more than one task can be placed at a PU. This constraint is valid only for tasks belonging to the same dependency group. For the example shown in Figure~\ref{fig:map}(b), \textit{task 0} is separated from \textit{task 1} and \textit{task 2} with a red dashed line. Thus, we couple only qubits $(q_2 \cdot q_4)$ and $(q_3 \cdot q_5)$ with a high penalty cost to prevent placing them on the same PU, which would otherwise be a valid solution for the solver. The first assignment constraint that says that a task can not be placed on multiple PUs at the same time remains unchanged.

\textit{Level adjustments.}
When the number of parallel tasks exceeds the number of available computing resources, an important decision has to be taken to prioritize a set of tasks in the most efficient way. This decision is reflected in the qubit matrix, i.e. the order of columns associated to specific tasks and corresponding
assignment constrain couplings. Multiple approaches exist in the field, but this study is out of the scope of this paper. Here, we apply a simple cut based on the task ID increment. Figure \ref{fig:map}(b) illustrates the case in which \textit{task 4} belongs to dependency level 1, but is moved to the next level. In case there are no available slots in the following group of tasks, an additional level is created.

\label{subsec:sub}

\begin{figure*}[htp]
\includegraphics[width=\linewidth]{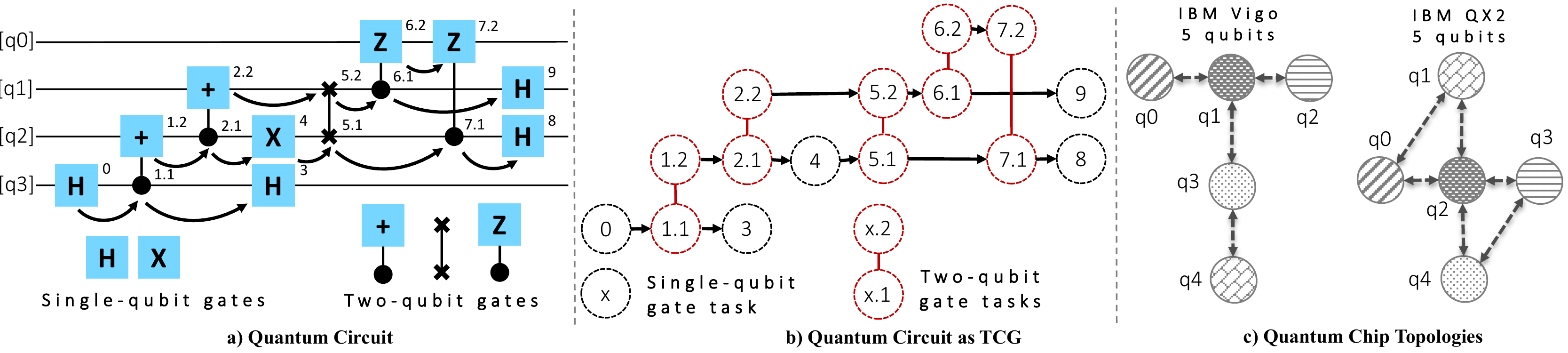}
\caption{Quantum circuit graph: gate-to-qubit assignment.}
\label{fig:qc}
\end{figure*}

\subsection{Domain-specific TCG partition}

Given the number of logical qubits together with the potential number of couplings and constrains per single problem, we quickly exhaust the physical capabilities of quantum machines. Therefore, an intelligent problem partition is required. There has been extensive research on graph partitioning~\cite{Graph_partitioning}. In this context, we apply the method shown in Figure~\ref{fig:map}(c). This method divides a TCG into sub-graphs (SGs) based on dependency levels. The example shown in Figure \ref{fig:map}(c) illustrates partitioning with two and three dependency levels per sub-QUBO1/2 and sub-QUBO3 respectively. The lowest degree of granularity corresponds to one dependency level per sub-QUBO. Further division of the problem will distort the concept of optimal parallel tasks assignment. 
The weakness of such a partitioning is that only communication edges inside a SG are regarded. Thus, multiple communication edges get excluded from the problem and are not represented in the qubit matrix. Excluded edges are labelled with red crosses in Figure~\ref{fig:map}(c). This may have a significant impact on the quality of the provided solution, especially for communication-intensive applications.

Part of the novelty of our work is improving the partition by applying an interactive previous-placement-dependent approach. This approach takes advantage of dependency level-based partitioning. Sub-QUBOs are solved one after another and each previous SG placement is used to enhance following sub-QUBOs. Our mapper extends the qubit matrix with additional virtual qubits--one per each unique source task of all excluded input edges (edges that are inputs to a SG). This qubit is associated with a specific PU because the previous task placement is already known at this point. 
In Figure~\ref{fig:map}(c), virtual qubits are shown as red crosses inside the sub-QUBO matrices and missed edges previously shown as crossed out are illustrated with red arrows. 

Our approach guides the solver towards a better solution than is possible with heuristics alone, but does not guarantee 
an optimal solution because the output edges of the sub-graphs are still excluded from the problem and the future placement is not available at this point.
It should also be emphasized that QUBO minimizes the sum of given costs, which are abstract positive numbers.
Minimizing the sum does not guarantee that parallel execution time is also minimized, if that is determined by the slowest task.

\subsection{Binary solution interpretation}

Figure \ref{fig:map2} illustrates the binary solution interpretation by mapping the example graph from Figure \ref{fig:map}(c) on the four-unit mesh architecture. Each block corresponds to a dependency level of the task-communication graph. It contains three illustrative components, i.e. a qubit sub-matrix with solution values, computation task placement corresponding to the solution and communication traffic based on the prior task placements. In case both source and destination tasks are placed on the same unit, the communication edge is marked as local communication. Local communications do not contribute to the data movement component of the objective function and represent the most favourable assignment for communication cost minimization.

\subsection{Computation and Communication costs}
\label{subsec:ccc}
Computation and communication costs have been previously discussed as abstract positive numbers. However, the nature of the cost metric determines whether the proposed method provides an optimal solution. If the cost is based on delay and the goal of task assignment is to minimize time, QUBO minimization will not provide the optimal placement. This is because QUBO minimizes the sum of the placement costs in each SG and it does not guarantee that if placed in parallel task execution time is minimum. For other metrics, such as data movement, power consumption, energy, the proposed method provides an optimal solution. 
{ 
\section{Gate Assignment Mapping Algorithm}
\label{sec:qmap}

\subsection{Quantum Circuits}
In the context of gate-based quantum computing, quantum algorithms are usually represented in the form of so called \textit{quantum circuits}. Figure \ref{fig:qc}(a) shows an example of the quantum circuit.
To avoid confusion, the qubits represented on the circuit will be referred to as \textit{logical} qubits and the real qubits inside a quantum computer as \textit{physical} qubits. Four horizontal lines represent logical qubit state evolution over time (from left to right). Single- and two-qubit gates are applied on specific qubits according to algorithm computations. Quantum circuits can be transformed into a task-communication graph similar to the classical algorithm transformation. In this case, quantum gates represent tasks that have dependencies (black arrows). Figure \ref{fig:qc}(b) shows the Quantum Circuit Graph (QCG) in the form of the TCG. A two-qubit gate becomes two connected tasks in the QCG. Moreover, two-qubit gates are directional, i.e. there are source and destination qubits in the pair. 

Topology-aware quantum gate assignment is based on physical qubit connectivity inside the quantum chip. Figure \ref{fig:qc}(c) shows an example of the 5-qubit chip connectivity. Arrows show not only the connection between two physical qubits, but also the supported direction for the two-qubit gates.
Because of the limited connectivity between qubits, not all two-qubit gates can be directly applied. For example, consider a circuit where a two-qubit gate is applied to logical qubits 0 and 3, and the circuit is matched to the architecture on Figure \ref{fig:qc}(c). There are two ways to map the qubits to circuit. First is to map the logical qubits to physical in a different order such that logical 0 and 3 are mapped to physical 0 and 2. Another is to swap the underlying logical qubit states, in case if they are already mapped to the architecture in the same order. For instance, if the states of qubits 2 and 3 are swapped, the physical qubit 2 now would contain the state of the logical qubit 3, making it possible to apply the desired 2-qubit gate.

\begin{figure*}[htp]
\includegraphics[width=\linewidth]{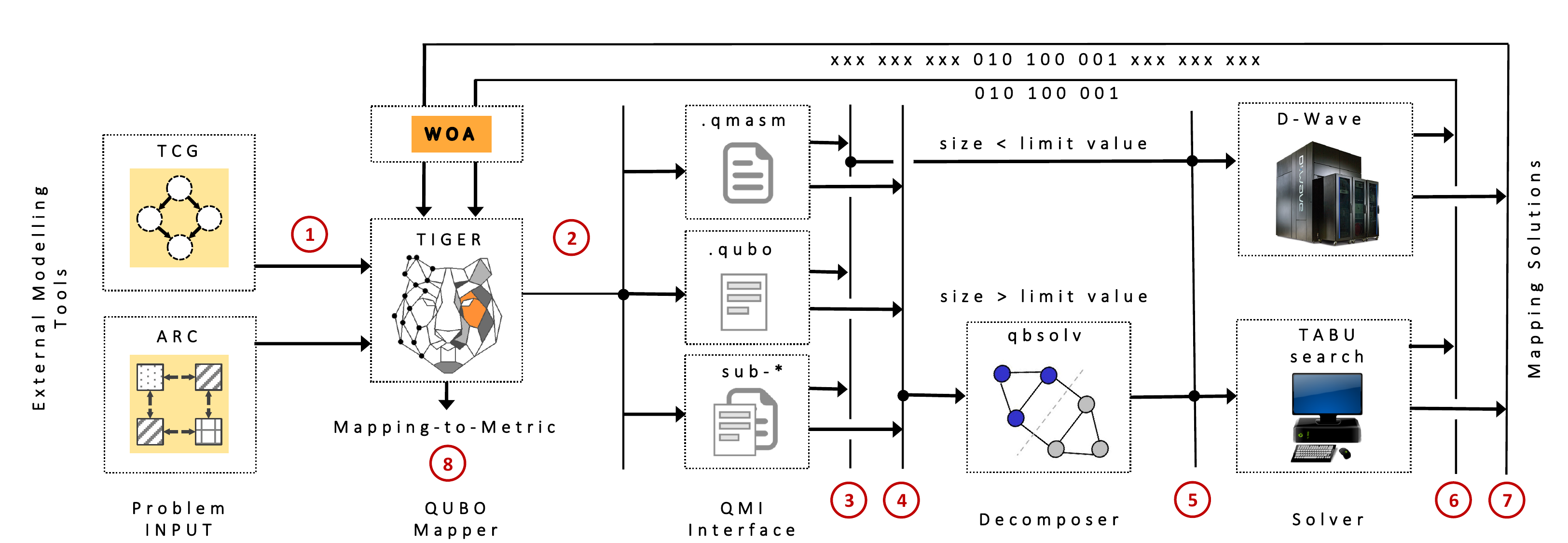}
\caption{Topology-aware task assignment using TIGER and quantum annealing.}
\label{fig:flow}
\end{figure*}

\subsection{Fidelity and SWAP operation costs}
Unlike a classical assignment optimization problem that minimizes computation and communication costs (described in Section \ref{subsec:ccc}), in quantum gate assignment optimization we target different metrics. One of the most important parameters for quantum computations in the NISQ era is \textit{fidelity}. Circuit fidelity is a measure of how much quantum information is preserved \cite{2018arXiv180710749M}. Due to the noise, the experimentally-obtained output qubit state is different from the desired output qubit state which would have been obtained in the ideal scenario. There is a direct correlation between the number of gates and circuit fidelity. 

Typically, in case of superconducting technology, single-qubit gates have higher fidelity than two-qubit gates, which require significantly more effort to tune and improve. Each physical qubit is unique in its properties and has different fidelity per gate. The fidelity resulting from mapping logical qubits and their corresponding gates to the underlying architecture's physical qubits will be referred to as \textit{fidelity\textsubscript{mapping}}. 

There are several types of two-qubit gates. \textit{SWAP} gate swap the states between two-qubits. A SWAP gate is usually decomposed into a sequence of three \textit{CNOT} two-qubit gates. CNOT belongs to the so-called \textit{native} set of gates that is supported by the control hardware and quantum chip technology. The need of this operation is dictated by the nature of quantum computation - it is not possible to \textit{make a copy} of a qubit state (\textit{no-cloning theorem} \cite{JPark} \cite{Wootters1982Single}). A SWAP gate is used to move the qubit state to the right location. Thus, the number of SWAP operations  \textit{N\textsubscript{swaps}} is similar to the data movement (communication) cost of the classical TCG. Consequently, the quantum state movement is required to satisfy chip connectivity. This movement comes at a cost, because two-qubit gates are the main source of infidelity in quantum circuits. The reduction in fidelity resulting from insertion of SWAP gates, each having fidelity \textit{fidelity\textsubscript{swap}}, will be referred to as \textit{fidelity\textsubscript{movement}}.

\begin{equation}  
\begin{split}
  &fidelity\textsubscript{movement}=(fidelity\textsubscript{swap})^{N_{swaps}} \\
  &fidelity\textsubscript{total}=fidelity\textsubscript{mapping}*fidelity\textsubscript{movement} \\
\end{split}
\label{eq:2}
\end{equation}

Since two-qubit gates have lower fidelity, quantum gate assignment optimization can be formulated as \textit{ N\textsubscript{swaps}} minimization. However, in order to obtain the best total fidelity for the quantum circuit both of the optimization parameters need to be taken into account, i.e. gate mapping fidelity and minimum number of SWAPs. That makes the optimization problem almost identical to the classical topology-aware task assignment on extremely heterogeneous architectures, where \textit{fidelity\textsubscript{mapping}} represents computation performance to be maximized and where  \textit{N\textsubscript{swaps}} represents the communication cost to be minimized. Equation~\ref{eq:2} shows how optimization of these two metrics can be reformulated as total fidelity \textit{fidelity\textsubscript{total} maximization}. A large number of recent studies target the total circuit fidelity maximization \cite{2018arXiv181000129D}. However, they solve the optimization problem of the circuit gate decomposition and assignment to minimize the number of gates, especially \textit{SWAP} gates, without consideration of \textit{fidelity\textsubscript{mapping}}.

\subsection{Weight Optimization Algorithm}
Ising machine weights allow us to vary the priority of one or another optimization metric. By scaling the weights associated with SWAP minimization, either the qubit fidelity or SWAP reduction can be prioritized. To scale the weights, a priority coefficient \textit{pref} is introduced.
To arrive at the optimal solutions either in terms of the resulting number of SWAP gates inserted or gate fidelity, we propose an optimization algorithm. It searches for the coefficient value that maximizes  fidelity\textsubscript{total}. Since fidelity\textsubscript{total} is obtained from fidelity\textsubscript{mapping} and fidelity\textsubscript{movement}, the algorithm can also find a solution with maximum fidelity\textsubscript{mapping} or minimum qubit movement. Due to infidelity of SWAP gates, a solution with minimum N\textsubscript{swaps} should correspond to maximum fidelity\textsubscript{total} solution. However, in a hypothetical fully-connected architecture where qubit movement constraint is eliminated, fidelity\textsubscript{mapping} would correspond to fidelity\textsubscript{total}. In such a scenario it would be practical to maximize only mapping fidelity. Optimizing only fidelity\textsubscript{mapping} or N\textsubscript{swaps} metric can also give an estimate of the bounds of these metrics in case if no optimal solution is known beforehand. Moreover, the proposed optimization algorithm can be suitable when it is needed to maintain a specific computation to communication ratio in task assignment, for example. The pseudocode is given in Algorithm \vref{algo}. The search starts with an initial preference coefficient, gets the corresponding metric value, for example fidelity\textsubscript{total}, and compares it to other solutions with a larger and smaller coefficient. The search space range is defined by setting the parameter \textit{sSpr}. How fast the algorithm converges is defined by the parameter \textit{sRed}, which reduces the search space at every step. For better local search space exploitation lines 6-17 can be repeated with $sSpr = \sqrt{sSpr}$.

\begin{algorithm}
\DontPrintSemicolon
\SetAlgoLined
\KwData{$QCG, ARC$} 
\KwResult{$fidelity_{best}, pref_{best}$}
 $sSpr = 2$
 \tcp{search spread, sets the search space range}
 $sRed = 0.9$
 \tcp{spread reduction, reduces $sSpr$ at every step for convergence}
 $pref_{best} = 0.05$
 \tcp{initial preference coefficient}
 $fidelity_{best} = tiger(QCG, ARC, pref)$\;
 \While{$sSpr$ $>$ 1}{
  $pref_{left} = pref/sSpr$\;
  $pref_{right} = pref*sSpr$\;
  $fidelity_{left} = tiger(QCG, ARC, pref_{left})$\;
  $fidelity_{right} = tiger(QCG, ARC, pref_{right})$\;
  \If{$fidelity_{left} > fidelity_{best}$}{
   $fidelity_{best} = fidelity_{left}$\;
   $pref_{best} = pref_{left}$\;
   }
  \If{$fidelity_{right} > fidelity_{best}$}{
   $fidelity_{best} = fidelity_{right}$\;
   $pref_{best} = pref_{right}$\;
   }
   $sSpr = sSpr * sRed$\;
 }
 \caption{Preference coefficient optimization}\label{algo}
\end{algorithm}

}

\section{TIGER}
\label{sec:tool}

\subsection{D-Wave programming environment}
Qbsolv~\cite{qbsolv} is an open source decomposing solver that focuses on large-scale problems that do not fit into physical hardware. In addition to the D-Wave annealer interface, qbsolv has an embedded classical solver that implements the tabu search algorithm~\cite{Glover:1997:TS:549765} to minimize the QUBO problem. Qmasm~\cite{7761637} is a quantum macro assembler that provides extra flexibility for programming. 
A qmasm program can be run on both D-Wave annealers and qbsolv classical solvers.

\subsection{TIGER tool flow} 

Figure~\ref{fig:flow} shows the tool flow for the task/gate assignment problem optimization. The key component of the flow is our proposed TIGER tool.
TIGER is an open-source QUBO mapper written in Python. It uses NetworkX python package~\cite{networkx} to create and manipulate TCG/QCG and ARC structures, i.e. computing the computation and communication costs for classical problems and fidelity and SWAP costs for quantum problems taking into account hardware (architecture) topology. We demonstrate TIGER on the D-Wave machine.

\begin{figure}[htp]
\centering
\includegraphics[width=\linewidth]{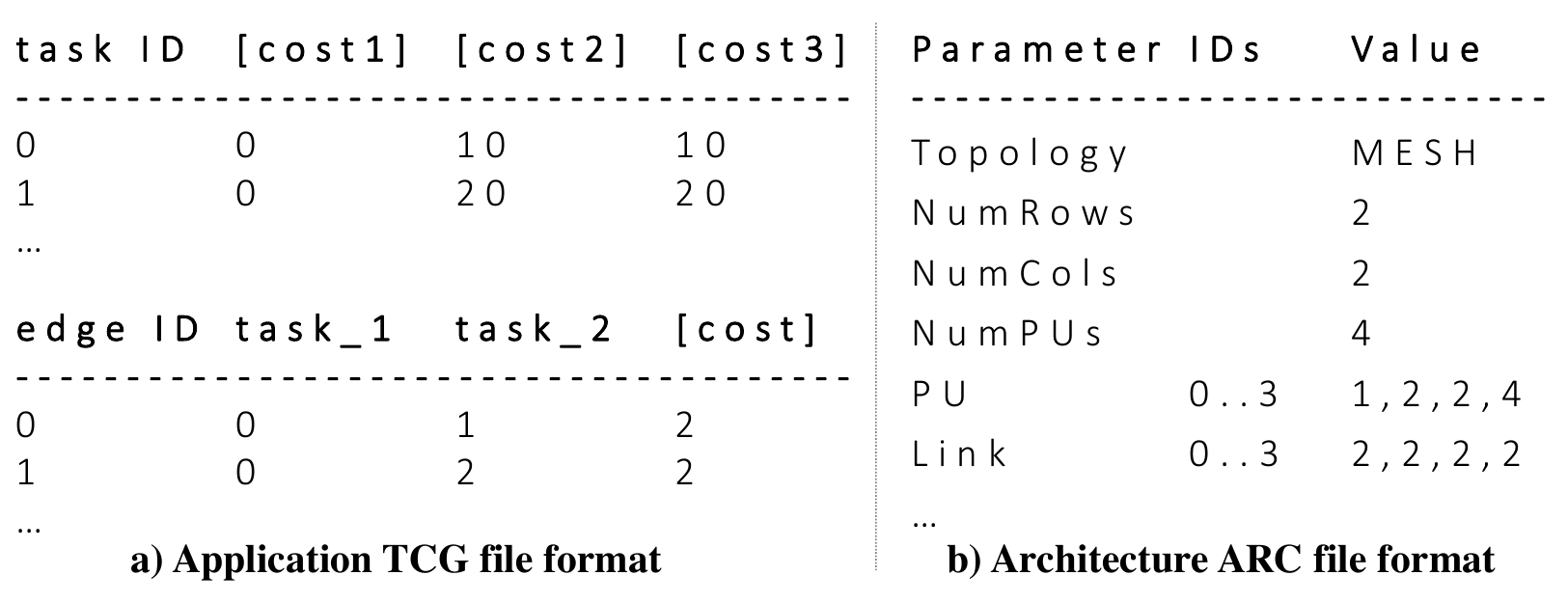}
\caption{Topology-aware task assignment problem input.}
\label{fig:format}
\end{figure}

TIGER receives two files as inputs (marked as red `1' to denote step 1), namely TCG or QCG and ARC (architecture).
TCG describes the classical application's TCG, QCG describes the quantum algorithm's QCG, while ARC describes the architecture (hardware topology).
The format of these files is presented in Figure~\ref{fig:format} (a) and (b). The TCG file consists of lines of two types associated to application tasks and edges. Task lines contain a task ID and multiple cost values each of a different type, e.g. number of integer, floating point, memory access instructions. Edge lines contain an edge ID, source and destination task IDs, and a cost value, e.g. the amount of data to be transferred between two tasks in bytes. The architecture file describes the architecture topology and its details such as number of rows and columns, number of PUs, and the capabilities of each PU and link such as cost per type of instructions, link throughput, etc.

Using the algorithm described in Section~\ref{sec:map}, TIGER maps input TCG and ARC files into the QUBO format and generates the QMI interface file (step `2'). It supports both qmasm and qubo formats and can generate a single file per problem or multiple files in case the QUBO partitioning option is chosen. If the size of the problem is less than the physical limit value, i.e. qubit sub-matrix size, 
QUBO or sub-QUBO can be directly solved (step `3'). Otherwise, it has to be further decomposed by qbsolv and then solved (step `4'). In both cases the problem is solved by two available solvers: the D-Wave annealer or a TABU search qbsolv implementation (step `5').

Finally, the solver generates mapping solutions that are sent back to the TIGER tool. If the solution corresponds to sub-QUBO (step `7'), it is used by TIGER to generate the next sub-QUBO as described in Section~\ref{subsec:sub}. If the solution is complete (step `6') or the last sub-QUBO problem is solved, TIGER calculates the final cost of the assignment through its Mapping-to-Metric (MtoM) interface (step `8'). This cost can be used to estimate the quality of the solution.  

\section{Results}
\label{sec:res}


\begin{table}[t]
\centering
\small
\caption{Benchmark suite}
\begin{tabular}{l|c|c|c}
\textbf{Workload} & \textbf{Problem size} & \textbf{Tasks \#} & \textbf{Edges \#} \\ \hline \hline
Ultrasound & 9x5x10 & 15 & 15 \\
RS-encoder & 32x28x8 & 141 & 140 \\
RS-decoder & 32x28x8 & 526 & 789 \\
\hline \hline
\end{tabular}
\label{tab:bench}
\end{table}

\subsection{Experimental setup}

Experiments are conducted on a hybrid classical-quantum system that consists of an Intel Core i7 running at 3.3 GHz with 16 GB 2133 MHz LPDDR3 and a D-Wave 2X (DW2X) quantum annealer~\cite{dwave} that has 1152 qubits and 2400 couplers. 

\begin{figure*}[htp]
\centering
\includegraphics[width=\linewidth]{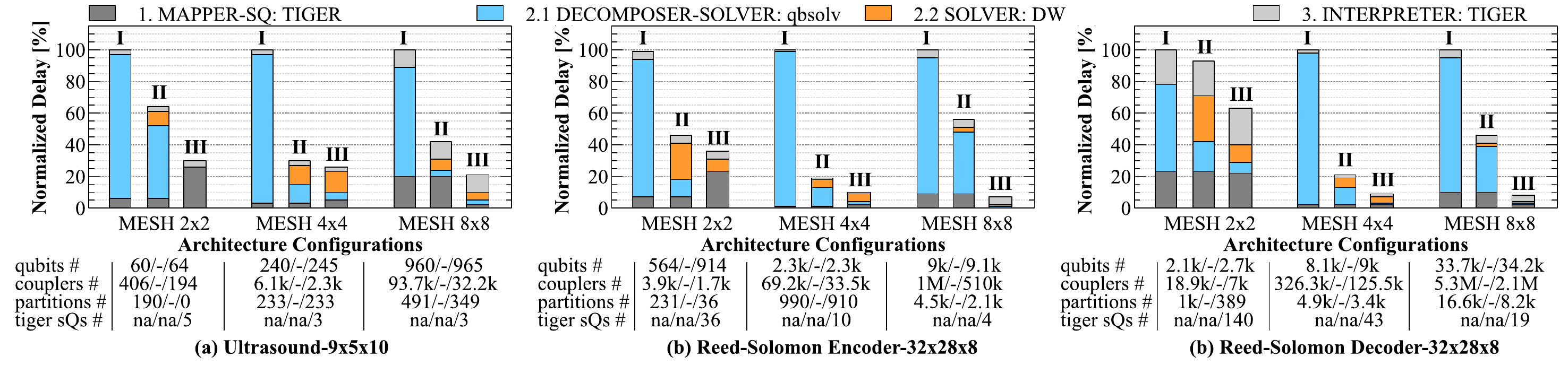}
\caption{Delay-to-Solution evaluation: \textbf{(I)} - classical TABU-search solver w/o TIGER sQ partition, \textbf{(II)} - quantum DW solver w/o TIGER sQ partition and \textbf{(III)} - quantum DW solver with TIGER sQ partition.}
\label{fig:exp1}
\end{figure*}


For classical TCG assignment optimization, we use three workload TCGs from the COSMIC benchmark set~\cite{6903410}. The choice of these three workloads is dictated by the differences in its problem size, number of tasks, and number of edges. A detailed analysis and classification of the application graphs in the context of the Ising model evaluation can provide additional insights. Such as study is out of the scope of this paper. The TCG files are provided by external modelling tool, i.e. the COSMIC benchmark suite~\cite{6903410}. Table \ref{tab:bench} shows the set of chosen benchmarks and their characteristics. 


For quantum QCG assignment optimization, we create the QCG files formatted for TIGER from the quantum benchmark suite \cite{zulehner2018efficient}. We create ARC files based on two IBM quantum chips \cite{ibmq}: \textit{IBM Yorktown (QX2)} with 5 qubits and \textit{IBM Vigo} with 5 qubits. Figure \ref{fig:qc} (c) illustrates these two topologies. The quantum benchmark suite~\cite{zulehner2018efficient} provides 48 circuits for 5-qubit chips. We reduce the circuit size down to 50 gate. 




\subsection{Tool flow evaluation}

For each workload we evaluate three scenarios:
\textit{(I)} TIGER QUBO mapper - qbsolv decomposer/TABU-search qbsolv solver - TIGER MtoM interpretor, \textit{(II)} TIGER QUBO mapper - qbsolv decomposer/DW solver - TIGER MtoM interpretor and \textit{(III)} TIGER QUBO mapper/TIGER SG partitioner - qbsolv decomposer/ TABU-search qbsolv solver - TIGER MtoM interpretor.
For each scenario, we vary the size of the architecture to a 2$\times$2 PU mesh, 4$\times$4 PU mesh, and an 8$\times$8 PU mesh.

Figure~\ref{fig:exp1} shows evaluation results. Here, we report the delay normalized to the total delay of the longest case. Each delay is also broken down to its four major components. In all cases, the longest scenario is the one fully executed on a classical computer, e.g. scenario \textit{I}. In addition, we show the number of logical qubits and couplers generated by TIGER's mapper (\textit{qubits \#} and \textit{couplers \#}), the number of partitions provided by qbsolv's decomposer (\textit{partitions \#}), and the number of SGs generated by TIGER's partitioner (\textit{tiger sQs \#}). The number of qubits in scenarios \textit{I} and \textit{II} is equal, but it is higher in scenario \textit{III} because additional qubits are required to define previous sub-QUBO placements as shown in Figure~\ref{fig:map}. Similarly, the number of couplers as well as the number of partitions in scenarios \textit{I} and \textit{II} are equal. It is lower in scenario \textit{III} due to the optimized QUBO mapping. The number of TIGER sub-QUBOs is reported only for scenario \textit{III}. In scenarios \textit{I} and \textit{II} this TIGER option is not applied (\textit{na}).

\begin{figure*}[htp]
\centering
    \begin{subfigure}{\linewidth}
        \includegraphics[width=\linewidth]{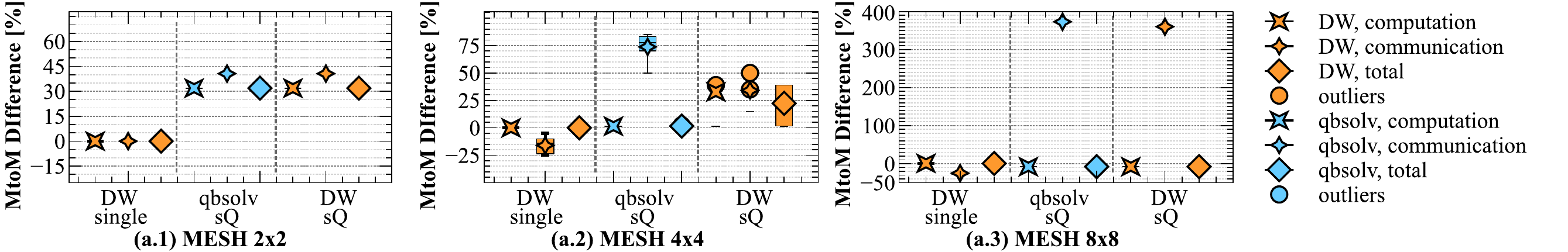}
        \caption{Ultrasound-9x5x10}
    \end{subfigure} 
    \begin{subfigure}{\linewidth}
        \includegraphics[width=\linewidth]{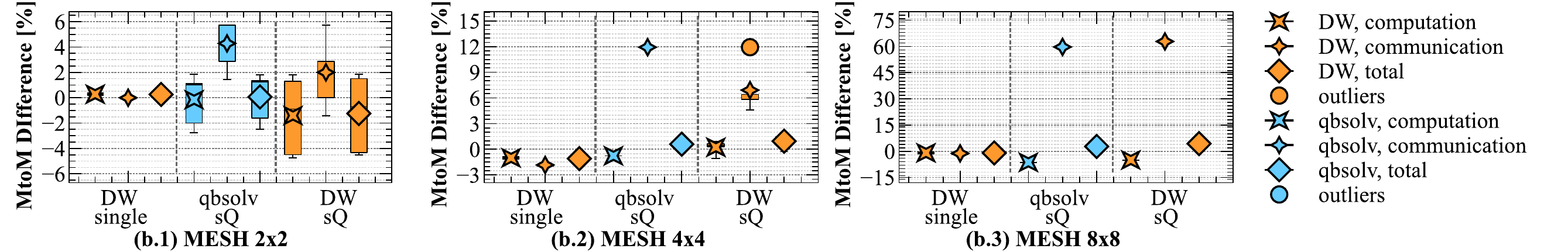}
        \caption{Reed-Solomon Encoder-32x28x8}
    \end{subfigure} 
    \begin{subfigure}{\linewidth}
        \includegraphics[width=\linewidth]{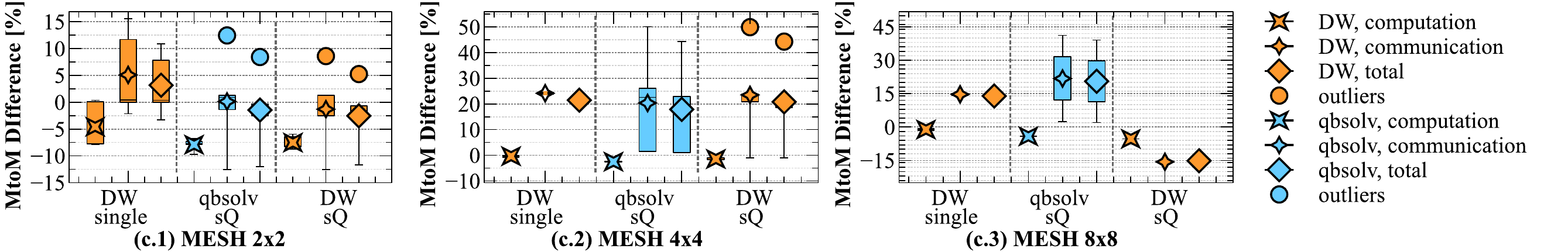}
        \caption{Reed-Solomon Decoder-32x28x8}
    \end{subfigure} 
\caption{Task assignment sensitivity and quality of the solution. \textbf{(DW, single)}: DW w/o sQ vs. classical TABU-search w/o sQ, \textbf{(qbsolv, sQ)}: classical TABU-search with sQ vs. classical TABU-search w/o sQ and \textbf{(DW, sQ)}: DW with sQ vs. classical TABU-search w/o sQ.}
\label{fig:exp2}
\end{figure*}

\textbf{Discussion:} Performance evaluation results prove that the physical quantum annealer, i.e. DW2X, can significantly reduce delay-to-solution compared to the classical qbsolv solver. For the given set of benchmarks and architecture configurations, the performance speedup of the \textit{DECOMPOSER-SOLVER} phase varies between 1.2$\times$ and 10.2$\times$. The major portion of this improvement is caused by the replacement of the classical solver with the quantum annealer. The average value of the DW2X access time is around 20ms. This time includes programming time, sampling time and post-processing time. The sampling phase consists of multiple sample batches, each of which includes annealing, readout, and additional delay that allows the quantum annealer to cool down to the initial state. The annealing time is 20\si{\us}. Although QUBO is solved by a physical quantum annealer, a significant amount of time associated to the problem decomposition is spent by \textit{qbsolv DECOMPOSER}. The total \textit{D-Wave SOLVER} phase is composed of multiple D-Wave accesses, where the number of accesses is determined by the number of partitioned calls provided by \textit{qbsolv DECOMPOSER}. Therefore, while using the quantum annealing solver the delay-to-solution phase highly depends on the quality of the classical decomposition.

In scenario \textit{III}, we evaluate the impact of the domain-specific partitioning integrated into the QUBO mapper, i.e. TIGER level partitioner.
Here, reported values represent the sum of all sub-QUBO parameters concerning the total number of qubits and couplers as well as delays per phase. Results show that by applying two-level QUBO partitioning (i.e. domain-specific first and classical qbsolv second), a massive speedup in time-to-solution can be achieved. For the given set of TCGs and ARCs, the \textit{DECOMPOSER-SOLVER} phase is reduced down to 6\% compared to the baseline scenario. Such an improvement has several sources. First, TIGER partition significantly simplifies the task for \textit{qbsolv DECOMPOSER}, which performs better on a smaller subset of qubits and coupler tasks than on a single large problem. Consequently, qbsolv generates fewer partition calls thereby reducing \textit{D-Wave SOLVER} phase delay. This effect is particularly noticeable for larger TCGs where the number of partitions is reduced twice. The total number of qubits and couplers is also different compared to the baseline. By applying the minimum number of qubits possible and adjusting the level of granularity (i.e. one sub-level per sub-QUBO), we reduce the total number of couplers. These improvements are achieved at the expense of having a larger number of qubits. This increase is 12\% by average compared to the baseline. On the other hand, additional partitioning can potentially impact the quality of the generated solution. This effect is evaluated in the following section.  

\subsection{Task assignment evaluation}

We evaluate the assignment quality and multiple-run sensitivity in three comparison scenarios: \textit{(i)} single QUBO on quantum annealer versus classical qbsolv solver (\textit{dw, single}), \textit{(ii)} partitioned sub-QUBOs versus single QUBO assignment on classical qbsolv solver (\textit{qbsolv, sQ}), and \textit{(iii)} partitioned sub-QUBOs on quantum annealer versus single QUBO assignment on classical qbsolv solver (\textit{DW, sQ}). Architecture configuration files represent a 2$\times$2, 4$\times$4, or 8$\times$8 heterogeneous PU MESH with an abstract PU acceleration factor varied from 1$\times$ to 4$\times$. Link cost is equal to 2. Figure~\ref{fig:exp2} shows the difference in computation, communication and total costs for the three evaluation scenarios compared to the baseline.

\textbf{Discussion:} In some cases, we obtain the same solution over multiple runs. If different solutions are returned, usually the variation is within 5\% from the mean value. For a given set of experiments, DW2X quantum solver provides solution improvements for a single QUBO compared to the classical TABU-search solver. Results show up to 8\% of computation cost improvement, up to 25\% of communication cost improvement, and up to 15\% of total improvement. Both qbsolv sQ and DW sQ scenarios show similar behaviour in most experiments. However, again DW2X quantum solver provides better solutions, e.g. RS-Encoder mapped on 2$\times$2 MESH and RS-Decoder mapped on 2$\times$2 MESH.

Qbsolv sQ and DW sQ evaluations show that dependency-level partitioning indeed can significantly impact assignment quality, namely its communication constituent. This impact increases when architecture size scales. MtoM communication difference rises from 35\% to 45\% and then to almost 4$\times$ for \textit{US} TCG mapped on 2$\times$2, 4$\times$4 and 8$\times$8 architectures shown in Figure~\ref{fig:exp2}(a). Similarly, it changes from -2\% to 6\% and then to 60\% for \textit{RS Encoder} TCG as shown in Figure~\ref{fig:exp2}(b). However, the computation constituent does not deteriorate. In both TCGs, task computation costs far outweigh communication edge cost. For instance, \textit{US} computation cost ranges between 4,510 and 3,461,112, while communication highest cost is 20, 60 and 140 for 2$\times$2, 4$\times$4, and 8$\times$8 MESHes respectively.
Thereby, calculated edge weights and associated qubits couplings have low impact on the total QUBO cost. Indeed, the total MtoM difference follows the computational cost behaviors, e.g. DW sQ in Figure~\ref{fig:exp2}(b.1) or qbsolv SQ in Figure~\ref{fig:exp2}(a.2). By prioritizing the edge cost versus task cost, the communication MtoM difference can be significantly reduced. 
In contrast, \textit{RS Decoder} TCG is communication intensive. The computation cost varies to up to 1,880, while the communication cost reaches 14,280 for 8$\times$8 MESH. In this case, DW sQ partition does not impact the solution quality, but improves it by up to 15\%. 


{
\subsection{Gate assignment evaluation}

\begin{figure}[t]
\centering
\includegraphics[width=\linewidth]{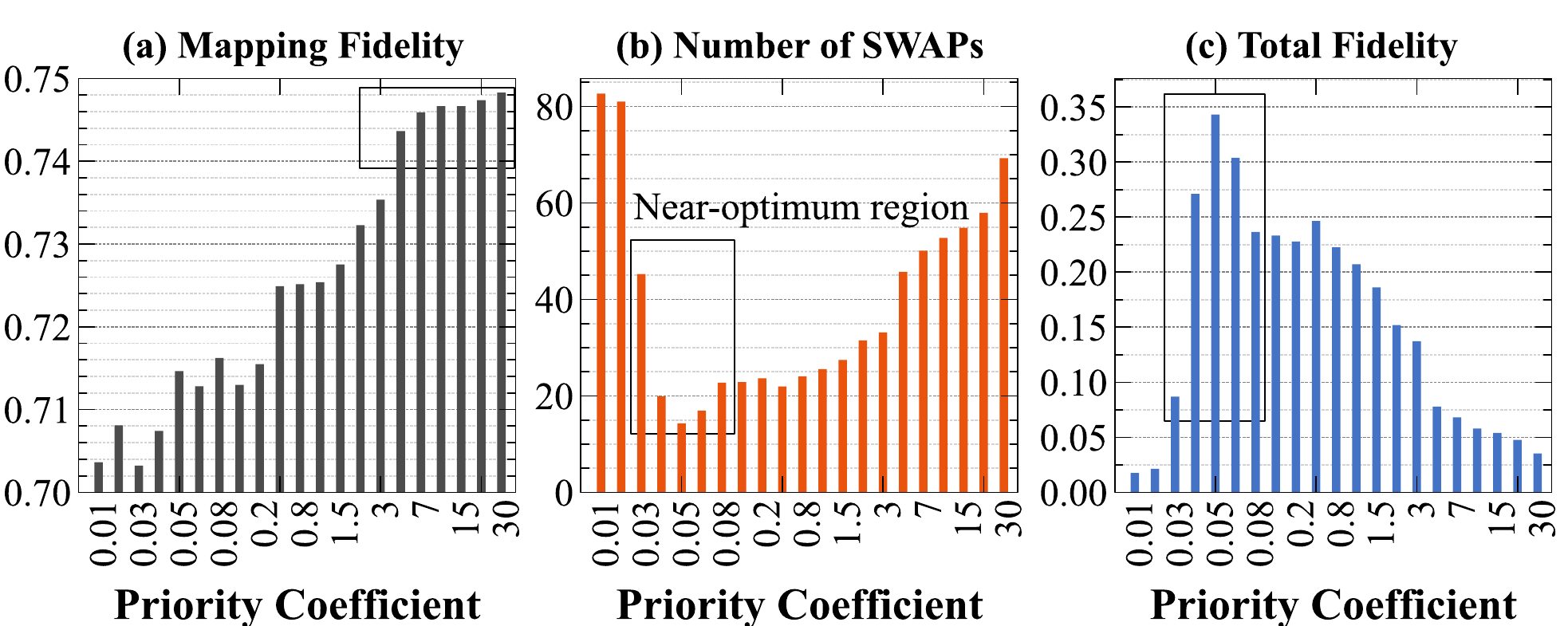}
\caption{IBM Vigo: mapping fidelity, number of swaps and total fidelity.}
\label{fig:vigo-all-coef}
\end{figure}

\begin{figure*}[htp]
\centering
    \begin{subfigure}[t]{0.49\textwidth}
        \includegraphics[width=\linewidth]{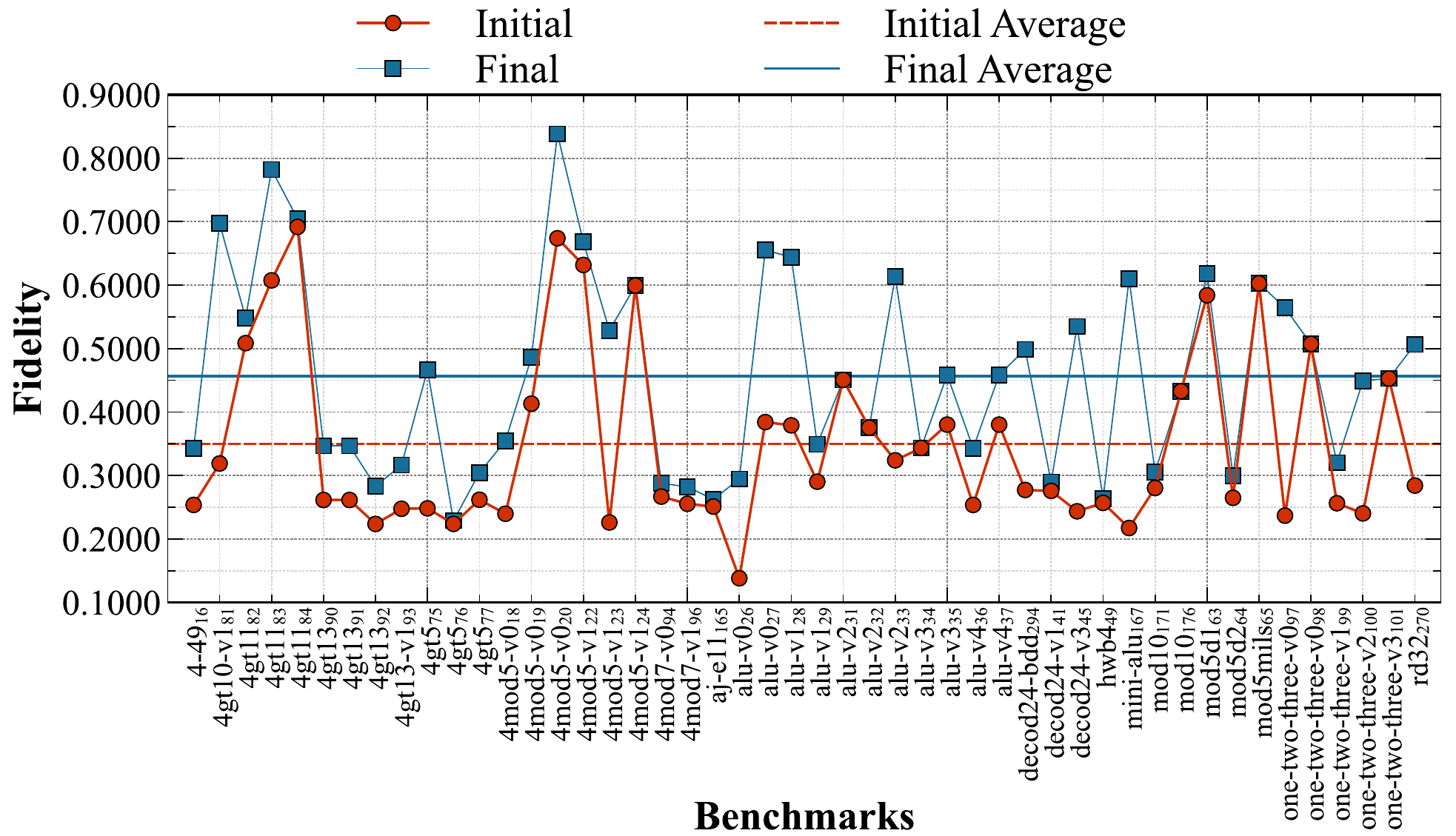}
        \caption{Vigo: Fidelity}
    \end{subfigure} 
    \begin{subfigure}[t]{0.49\textwidth}
        \includegraphics[width=\linewidth]{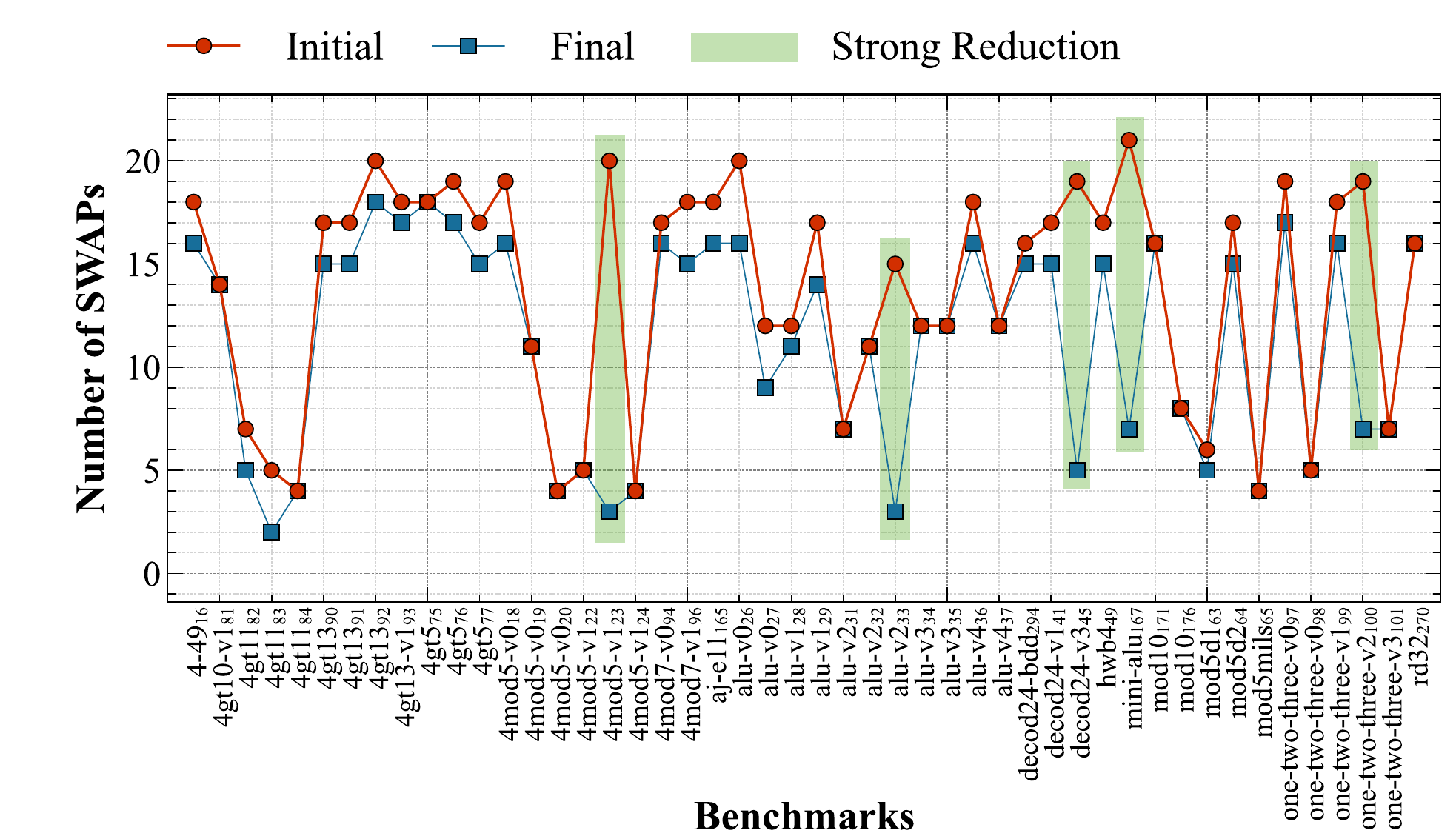}
        \caption{Vigo: Number of SWAPs}
    \end{subfigure} 
    \begin{subfigure}[t]{0.49\textwidth}
        \includegraphics[width=\linewidth]{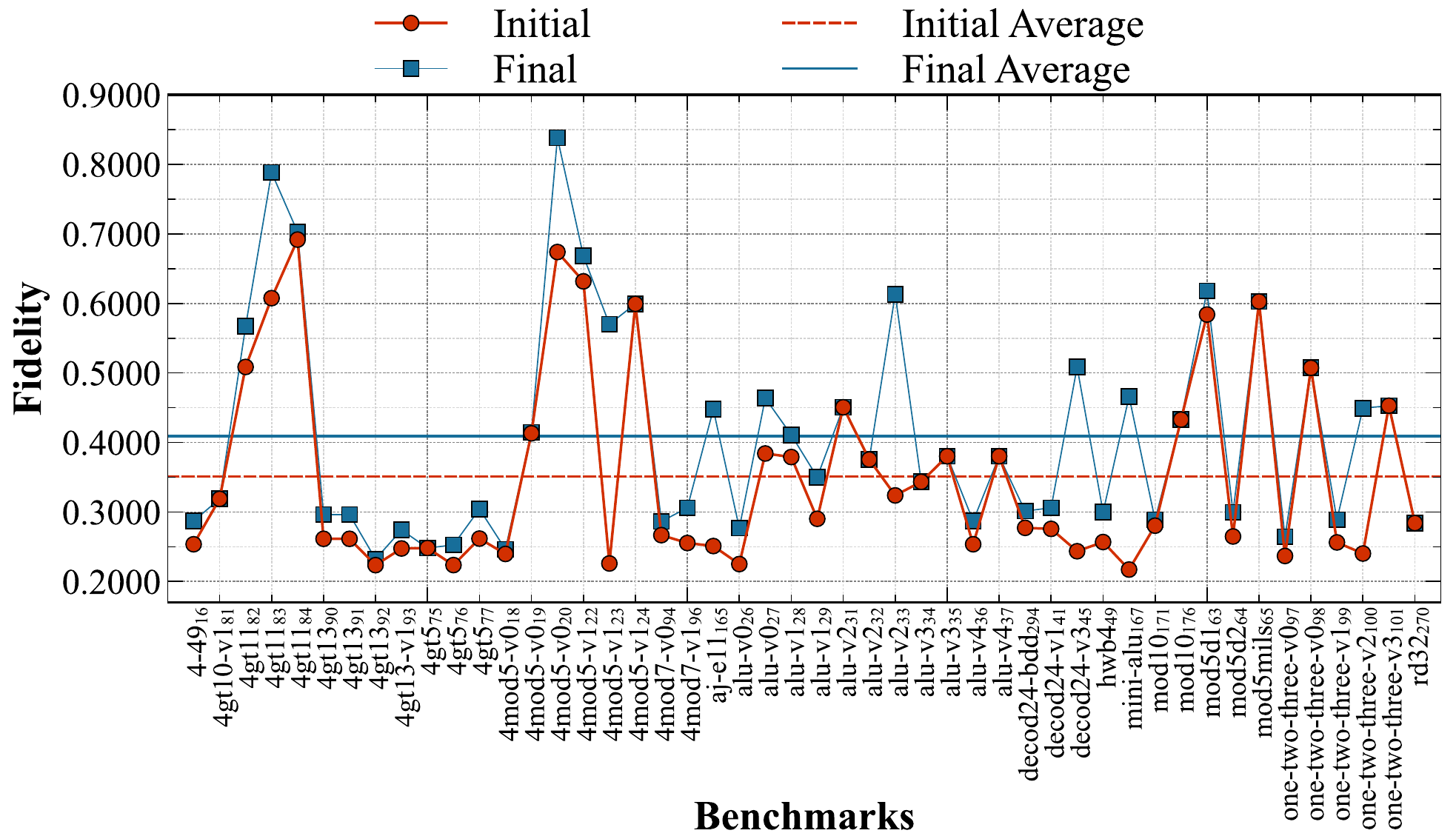}
        \caption{QX2: Fidelity}
    \end{subfigure} 
    \begin{subfigure}[t]{0.49\textwidth}
        \includegraphics[width=\linewidth]{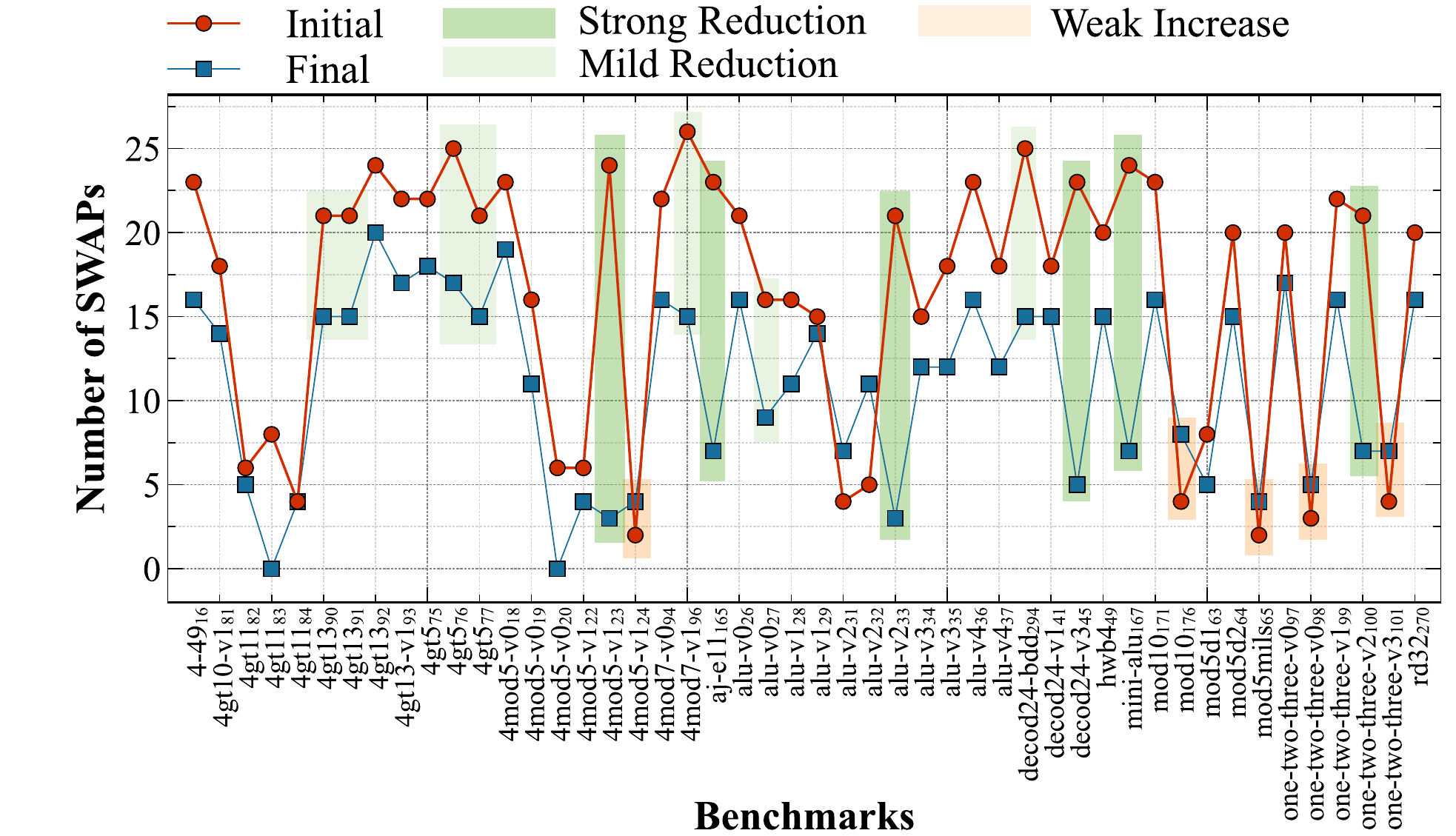}
        \caption{QX2: Number of SWAPs}
    \end{subfigure}
\caption{Quantum gate assignment: wieght optimization algorithm search}
\label{fig:qexp}
\end{figure*}

In order to make an estimation of how scaling the weights would correlate with N\textsubscript{swaps} and fidelity\textsubscript{mapping} a set of experiments was performed on the QCGs mentioned in Section 6.1.2. The preference coefficient varies from 0.01 to 30. Figure \ref{fig:vigo-all-coef} shows the mapping fidelity (fidelity\textsubscript{mapping}), number of swaps (N\textsubscript{swaps}) and total fidelity for different coefficient values. Smaller coefficients minimize qubit state movement, while larger ones prioritize mapping fidelity instead. Black box shows a near-optimum region of the priority coefficient. Using the priority coefficient smaller than 0.05 results in invalid solutions being produced by the algorithm and can even lead to the opposite effect, increasing N\textsubscript{swaps} instead. Setting the coefficient larger than 20 provides only small improvement of fidelity\textsubscript{mapping}, but it only happens in some architectures and incurs an inadequate number of additional SWAPs. Hence, applicable coefficient values that produce the minimum N\textsubscript{swaps} and maximum fidelity\textsubscript{mapping} are approximately 0.05 and 20, respectively. Total fidelity strongly correlates with the number of SWAPs and mapping fidelity plays a negligible role in this scenario.

\textbf{Discussion:} Since fidelity\textsubscript{movement} coming from N\textsubscript{swaps} has a larger impact on fidelity\textsubscript{total}, usually N\textsubscript{swaps} is minimized and gate fidelity is not considered. It means that the priority coefficient that maximizes fidelity\textsubscript{total} is the same that minimizes N\textsubscript{swaps}, i.e. 0.05. However, as connectivity in quantum computing architectures increases, qubit movement might become less significant. In such a context maximization of fidelity\textsubscript{total} would be entirely dependent on fidelity\textsubscript{mapping}. 

\subsubsection{Weight optimization algorithm evaluation}
\label{subsubsec:woae}
To tackle any possible scenario, fidelity\textsubscript{total} can be maximized regardless of connectivity and gate fidelity. The priority coefficient that allows such a maximization is unknown, and can vary for every different circuit and architecture. We study the proposed  weight optimization algorithm to assess its efficiency in finding the optimal priority coefficient for a combination of quantum circuit and device topology. 

Figure \ref{fig:qexp} shows total fidelity and number of SWAPs optimization using WOA algorithm for multiple circuits for IBM Vigo and IBM QX2 topologies. The results include initial value at the beginning of the algorithm execution and the final value.  For IBM Vigo topology (results in Figure \ref{fig:qexp} (a) and (b)), the WOA finds the priority coefficient that reduces the number of SWAPs from the initial step value in 62.5\% of cases. In 37.5\% of cases the number of SWAPs remains unchanged. The results with strong reduction are highlighted in green. In average, WOA improves total fidelity by 39\% for IBM Vigo topology. For IBM QX2 topology (results in Figure \ref{fig:qexp} (c) and (d)), the WOA finds the priority coefficient that reduces the number of SWAPs from the initial step value in 83.3\% of cases. In one case the number of SWAPs remains unchanged, and in 14.6\% of cases WOA provides weak increase of the SWAPs number. In average, WOA improves total fidelity by 107\% for IBM QX2 topology.

\textbf{Discussion:} The results show significant difference in WOA performance when applied on different topologies. While in general WOA allowed us finding more suitable combination of QUBO weights (preference coefficient) for both topologies, IBM QX2 mapping is much more sensitive towards priority coefficient choice. Moreover, in few cases WOA missed optimal solution that resulted in a weak increase in SWAPs number compared to the initial state value. We believe, that the reason lies in the complexity of the topology graph that calls for the QUBO weights adjustments to find the most suitable combination in a near-optimum region.

\begin{figure*}[htp]
\centering
\includegraphics[width=\linewidth, trim=50 360 50 50, clip ]{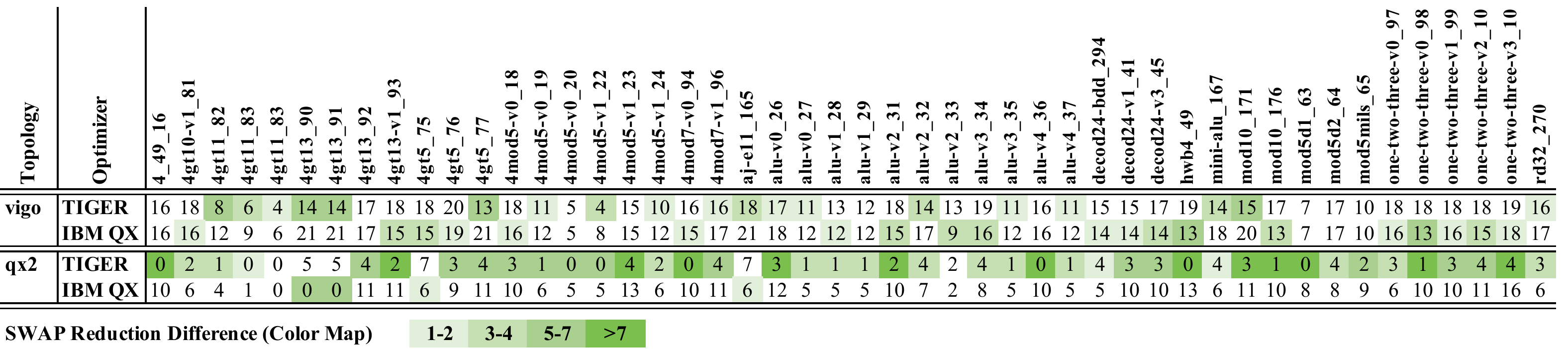}
\caption{Optimizer comparison: TIGER vs. IBM QX}
\label{fig:comp}
\end{figure*}

\subsubsection{Comparison}

Finally, we compare the performance of TIGER topology-aware SWAP optimizer against the IBM QX optimizer. Figure \ref{fig:comp} shows the comparison results across multiple circuits for two topologies, i.e. \textit{vigo} and \textit{qx2}. The numbers show the final number of SWAPS. The SWAP reduction color map highlights the cases when one of the optimizer provides a better result with the SWAP number differences as follow: (i) 1-2 SWAPs, (ii) 3-4 SWAPs, (iii) 5-7 SWAPs or (iv) more than 7. For the vigo topology, TIGER and IBM QX provides same SWAP number in 18.7\% of cases; IBM QX outperforms TIGER in 41.7\% of cases with the total reduction difference of 51 SWAPs; and TIGER outperforms IBM QX in 39.6\% of cases with the total reduction difference of 59 SWAPs. For the qx2 topology, TIGER and IBM QX provides same SWAP number only in 4.2\% of cases; IBM QX outperforms TIGER in 8.3\% of cases with the total reduction difference of 12 SWAPs; and TIGER significantly outperforms IBM QX in 87.5\% of cases with the total reduction difference of 260 SWAPs. Moreover, TIGER found the perfect mapping reducing the data movement to 0 SWAPs in 16.7\% of cases, while IBM QX found the perfect matching only in 4.2\% of cases.

\textbf{Discussion:} Similar to the WOA evaluation results (see section \ref{subsubsec:woae}), the comparison results show significant difference when applied on different topologies. TIGER allowed us significantly improve the mapping for IBM QX2 topology compared to the IBM QX optimizer. We believe, that the reason also lies in the topology graph complexity. Classical IBM QX optimizer is not suitable for more complex topologies with a larger number of potential combinations, while TIGER optimizer allows us to find the `perfect' mapping regardless.

}
\section{Conclusions}
\label{sec:con}

In this paper, we propose an algorithm for solving the topology-aware task/gate assignment problem on physical Ising machines in order to accelerate and improve the quality of the solution to this challenging NP-complete problem. We implement our solution in our TIGER tool that transforms weighted task-communication, quantum circuit, and architecture graphs into an appropriate format of the Hamiltonian function. Our solution takes into account both computation and communication costs for the classical problem or fidelity and SWAP number for the quantum problem. We evaluate the proposed approach using D-Wave's quantum annealer. In order to overcome existing physical limitations of current quantum annealers, we propose domain-specific partitioning based on the task-communication graph dependency levels. Also, we propose weight optimization algorithm that enables adjusting the model parameters and find better solutions. We integrate TIGER into the D-Wave software stack that enables us to apply both our proposed dependency-level partitioning as well as the partitioning provided by the qbsolv tool in a dynamic iterative way. We demonstrate that our method can reach 15\% higher-quality solutions 9\% faster compared to the classical qbsolv heuristic algorithm.
Finally, TIGER reduces the data movement cost by 68\% in average for quantum circuit assignment compared to the IBM QX optimizer~\cite{ibmq}.
Our work alleviates the concern that task mapping may hinder high-quality solutions on future quantum accelerators with more physical qubits and complex connectivity.
The TIGER tool is publicly available online \footnote{https://github.com/lbnlcomputerarch/tiger}.

For future work, we consider three major directions: 
\begin{itemize}
\item \textbf{Comparison to a wide range of classical scheduling tools}: we plan to design a methodology to compare the hardware optimizer, i.e. Ising machine, to existing heuristic software tools. 
\item \textbf{Use other Ising machines}: we plan to expand our study running the problem on other Ising machines, such as digital annealer \cite{fujitsu} and coherent Ising machine \cite{optical}. 
\item \textbf{Problem partitioning algorithms and additional constrains mapping}: we plan to evaluate additional graph partitioning algorithms and alternative problem mapping algorithms, e.g. assigning multiple tasks in one node based on the node capacity.
\end{itemize}

\begin{acknowledgements}
The research leading to these results has received funding from the the U.S. Department of Energy, grant agreement n\textsuperscript{o} DE-AC02-05CH11231. 
\end{acknowledgements}



\bibliographystyle{spmpsci}
\bibliography{cf19} 

\end{document}